# Phenomenological modeling of Geometric Metasurfaces

Weimin Ye[1,2*], Qinghua Guo[1,3*], Yuanjiang Xiang[3], Dianyuan Fan[3] and Shuang Zhang[1†]

[1]*School of Physics & Astronomy University of Birmingham Birmingham, B15 2TT , UK*
[2] *College of Optoelectronic Science and Engineering, National University of Defense Technology, Changsha, 410073, China*
[3]*SZU-NUS Collaborative Innovation Centre for Optoelectronic Science & Technology, College of Optoelectronic Engineering, Shenzhen University, Shenzhen 518060, China*

**Abstract**:

Metasurfaces, with their superior capability in manipulating the optical wavefront at the subwavelength scale and low manufacturing complexity, have shown great potential for planar photonics and novel optical devices. However, vector field simulation of metasurfaces is so far limited to periodic-structured metasurfaces containing a small number of meta-atoms in the unit cell by using full-wave numerical methods. Here, we propose a general phenomenological method to analytically model metasurfaces made up of arbitrarily distributed meta-atoms based on the assumption that the meta-atoms possess localized resonances with Lorentz-Drude forms, whose exact form can be retrieved from the full wave simulation of a single element. Applied to phase modulated geometric metasurfaces, our analytical results show good agreement with full-wave numerical simulations. The proposed theory provides an efficient method to model and design optical devices based on metasurfaces.

---

[*] These authors contribute equally to this work.
[†] s.zhang@bham.ac.uk

## I. Introduction

Metasurface is a thin layer of structured surface consisting of an array of planar metallic antennas with thickness far less than the wavelength of light. By tailoring the orientation and anisotropic resonances of meta-atoms, metasurfaces can be designed to control the phase, amplitude and polarization of light at the nanoscale, which has opened door for various planar photonic and optical devices with low manufacturing complexity [1-3]. They have been employed for ultra-thin wave plates [4-6], phase gradient plates [7-9], planar lenses [10, 11], optical vortex plates [7, 12] and holograms [13-15]. Among various types of metasurfaces, a special type of metasurfaces that manipulate the phase of light by the orientation of the antennas have shown great potential for practical applications due to the simplicity and robustness of the phase control. The underlying mechanism of the phase control is a Pancharatnam-Berry phase in the polarization state, which leads to strong spin–orbital interaction of light [16-19]. This concept has been used to realize spin controlled directional coupler for surface waves [20, 21] and broadband holograms [22, 23].

Up to date, theoretical modeling of metasurfaces is mainly limited to full-wave numerical simulations based on the Finite-Difference Time-Domain (FDTD) method and the Finite Element Method (FEM). Reported analytical modeling for metasurfaces, such as surface admittance method [6, 24-26] and homogenization with a Lorentz-formed polarizability [23], have only been applied to uniform metasurfaces with unit cell consisting of a single plasmonic element. So far, there has been no simple analytical modeling on the diffractive efficiencies of metasurfaces with complex phase profiles. In particular, for metasurfaces used as metalens, holograms and vortex plates that consist of a large number of different meta-atoms, there have been no efficient theoretical methods that could be used to accurately predict their optical performance. In this paper, we propose a general analytical theory to model geometric metasurfaces with arbitrarily distributed phase profiles. The method is shown to agree very well with, but much more efficient than the full-wave numerical simulations.

## II. Analytical Theory of Metasurface

Without loss of generality, we consider a metasurface sandwiched between the homogenous and isotropic cover and substrate, as depicted in Fig. 1. Where, $\boldsymbol{E}_\perp^{C\pm}$ ($\boldsymbol{E}_\perp^{S\pm}$) denote tangential component of electric fields propagating along $\pm z$ direction at the interface between the cover (substrate) and metasurface.

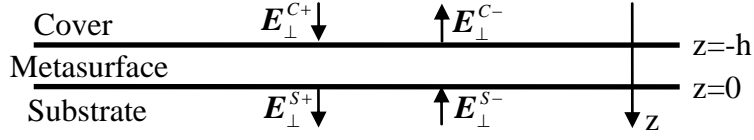

Figure 1. Schematic illustration of the model of metasurface.

Because the metasurface has a negligible thickness $h$ ($h \ll \lambda$), the overall tangential component of the electric fields is assumed to be continuous across the metasurface. Therefore the electric fields on the interfaces of the cover and substrate satisfy

$$\boldsymbol{E}_\perp^{C+} + \boldsymbol{E}_\perp^{C-} = \boldsymbol{E}_\perp^{S+} + \boldsymbol{E}_\perp^{S-} = \boldsymbol{E}_\perp \tag{1}$$

For the sake of simplicity, we only consider metasurfaces made up of achiral meta-atoms and neglect the surface current induced by the magnetic fields of light. The tangential components of magnetic fields at the interfaces of the cover and substrate thus satisfy

$$\boldsymbol{e}_z \times \left(\boldsymbol{H}_\perp^{S+} + \boldsymbol{H}_\perp^{S-} - \boldsymbol{H}_\perp^{C+} - \boldsymbol{H}_\perp^{C-}\right) = -i\omega\varepsilon_0 \boldsymbol{\chi} \cdot \boldsymbol{E}_\perp \tag{2}$$

Where, $\chi$ is the electric susceptibility tensor of the metasurface. Due to their deep subwavelength sizes, each antenna can be considered as an electric dipole located at $(x_m, y_n)$. The susceptibility tensor can be written as,

$$\chi(X) \approx \sum_{m,n} \delta(x - x_m) \delta(y - y_n) \boldsymbol{\alpha}_{mn} \tag{3}$$

In Eq. (3), $\boldsymbol{\alpha}_{mn}$ is the local polarizability tensor of the meta-atom located at $(x_m, y_n)$. Now we consider a geometric phase metasurface consisting of an array of anisotropic antennas of identical geometry but different orientations as schematically shown in Fig. 2 (a), where the Pancharatnam-Berry phase for circularly incident beam is determined by the orientation angle of each antenna.

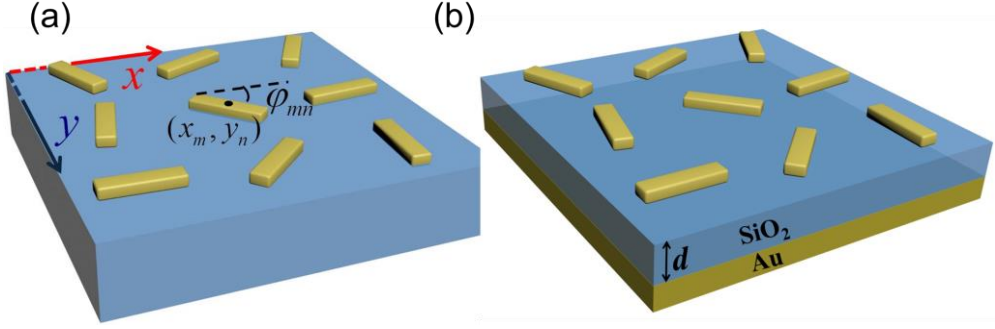

Figure 2. Schematic of a metasurface consisting of identical antennas with different orientations (a) arranged on an infinite substrate; (b) on top of SiO2/Au layers, where the thickness of SiO2 is $d$, and the thickness of Au is much greater than the skin depth such that no light can transmit through.

The local polarizability tensor can be expressed as

$$\boldsymbol{\alpha}_{mn} = \alpha_{mn+} + \alpha_{mn-} \begin{bmatrix} \cos 2\varphi_{mn} & \sin 2\varphi_{mn} \\ \sin 2\varphi_{mn} & -\cos 2\varphi_{mn} \end{bmatrix}; \quad \alpha_{mn\pm} = \frac{\alpha_{mn}^S \pm \alpha_{mn}^P}{2} \quad (4)$$

where $\varphi_{mn}$ is the antenna orientation angle, $\alpha_{mn}^S$ and $\alpha_{mn}^P$ are two orthogonal components of the local polarizability tensor of the anisotropic antenna. Note that when $\varphi_{mn}$ is 0, the polarizability is reduced to a diagonalized tensor [$\alpha_{mn}^S$, 0; 0  $\alpha_{mn}^P$]. The frequency responses of the two orthogonal components can be assumed to have Lorentz-Drude forms. That is

$$\alpha_{mn}^{S(P)}(\omega) = \frac{-cg_{mn}^{S(P)} \gamma_{mn}^{S(P)}}{\omega^2 - \omega_{mn}^{S(P)2} + i\omega\gamma_{mn}^{S(P)}} \quad (5)$$

where $g_{mn}^{S(P)}$ is a parameter indicating how strong the meta-atom couples with the incident electric fields, $c$ is the light speed in vacuum, $\omega_{mn}^{S(P)}$ and $\gamma_{mn}^{S(P)}$ are the resonance frequency and the damping rate of the meta-atom, respectively.

We can express the electric field and susceptibility tensor as the sum of various spatial frequency components through a Fourier expansion

$$\chi(X) = \sum_{k_x,k_y} e^{ik_x x + ik_y y} \bar{\boldsymbol{\alpha}}(k_x, k_y); \quad \bar{\boldsymbol{\alpha}}(k_x, k_y) = S^{-1} \sum_{m,n} e^{-ik_x x_m - ik_y y_n} \boldsymbol{\alpha}_{mn}$$

$$\boldsymbol{E}_\perp^{S(C)\pm}(X) = \sum_{k_x,k_y} e^{ik_x x + ik_y y} \boldsymbol{E}_{\perp k_x,k_y}^{S(C)\pm}; \quad \boldsymbol{E}_{\perp k_x,k_y}^{S(C)\pm} = \sum_{m,n} e^{-ik_x x_m - ik_y y_n} \boldsymbol{E}_{\perp mn}^{S(C)\pm} \quad (6)$$

where $S$ is the area of the metasurface. Combining Eq. (1-3) and Eq. (6), the Fourier expansions of electric and magnetic fields at the interfaces of the cover and substrate satisfy

$$E^{C+}_{\perp k_x,k_y} + E^{C-}_{\perp k_x,k_y} = E^{S+}_{\perp k_x,k_y} + E^{S-}_{\perp k_x,k_y}; \tag{7}$$

$$\boldsymbol{\eta}^S(k_x,k_y)\left(E^{S+}_{\perp k_x,k_y} - E^{S-}_{\perp k_x,k_y}\right) - \boldsymbol{\eta}^C(k_x,k_y)\left(E^{C+}_{\perp k_x,k_y} - E^{C-}_{\perp k_x,k_y}\right)$$
$$= (-i\omega\varepsilon_0)\sum_{k'_x,k'_y}\bar{\boldsymbol{\alpha}}(k_x - k'_x, k_y - k'_y)\cdot\left(E^{S+}_{\perp k'_x,k'_y} + E^{S-}_{\perp k'_x,k'_y}\right) \tag{8}$$

In Eq. (8), $\boldsymbol{\eta}^S(k_x,k_y)$ is the admittance tensor of the substrate, that relates the electric field and magnetic field in the media, $\begin{bmatrix}-H^{\pm S}_{yk_x,k_y}\\H^{\pm S}_{xk_x,k_y}\end{bmatrix} = \pm\boldsymbol{\eta}^S(k_x,k_y)\begin{bmatrix}E^{\pm S}_{xk_x,k_y}\\E^{\pm S}_{yk_x,k_y}\end{bmatrix}$ and can be expressed as

$$\boldsymbol{\eta}^S(k_x,k_y) = \frac{-\sqrt{\varepsilon^S/\mu^S}}{k_0 n_S\sqrt{k_0^2 n_S^2 - k_x^2 - k_y^2}}\begin{bmatrix}k_0^2 n_S^2 - k_y^2 & k_x k_y \\ k_x k_y & k_0^2 n_S^2 - k_x^2\end{bmatrix} \tag{9}$$

Where, $\varepsilon^S$, $\mu^S$ and $n_S$ are the permittivity, magnetic permeability and refractive index of the substrate. The expression for the admittance tensor of the cover is given by replacing 'S' by 'C' in Eq. 9.

It has been shown that the diffraction efficiency of metasurface can be dramatically enhanced through a three layer design, where the plasmonic antenna layer is on top of a homogeneous dielectric substrate layer of finite thickness $d$ and a thick ground metal layer such that the metausrface operate in reflection mode (Fig. 2b). It has been shown that, with suitably designed thickness of the substrate layer, the interplay between the antenna resonance and the Fabry-Perot resonance can lead to very high efficiency over a broad bandwidth [23]. We next extend our analytical analysis to such a multilayer design. Based on Eq. (7) and (8), the electric and magnetic fields at the interfaces of the cover and metasurface satisfy the following set of equations,

$$E^{C+}_{\perp k_x,k_y} + E^{C-}_{\perp k_x,k_y} = E^S_{\perp k_x,k_y} \tag{10}$$

$$\bar{\boldsymbol{\eta}}^S(k_x,k_y)\cdot E^S_{\perp k_x,k_y} - \boldsymbol{\eta}^C(k_x,k_y)\left(E^{C+}_{\perp k_x,k_y} - E^{C-}_{\perp k_x,k_y}\right) = (-i\omega\varepsilon_0)\sum_{k'_x,k'_y}\bar{\boldsymbol{\alpha}}(k_x - k'_x, k_y - k'_y)\cdot E^S_{\perp k'_x,k'_y} \tag{11}$$

where $\bar{\boldsymbol{\eta}}^S$ is an effective admittance that take into account the reflection by the metal ground plane and is expressed as,

$$\bar{\boldsymbol{\eta}}^S(k_x,k_y) = \boldsymbol{\eta}^S(k_x,k_y)\boldsymbol{R}^{-1}(k_x,k_y)\begin{bmatrix} -i\tan\psi^{\text{TE}}_{k_x,k_y} & 0 \\ 0 & -i\tan\psi^{\text{TM}}_{k_x,k_y} \end{bmatrix}\boldsymbol{R}(k_x,k_y) \quad (12)$$

where

$$\boldsymbol{R}(k_x,k_y) = \frac{1}{k_\perp k_z}\begin{bmatrix} k_y k_z & -k_x k_z \\ k_x k_0 n_S & k_y k_0 n_S \end{bmatrix}; \quad k_z = \sqrt{k_0^2 n_S^2 - k_x^2 - k_y^2}; \quad k_\perp = \sqrt{k_x^2 + k_y^2}$$

$$\psi^{\text{TE(TM)}}_{k_x,k_y} = k_z d - i\ln(\bar{r}^{\text{TE(TM)}}_{k_x,k_y})/2$$

In the above expressions, $\bar{r}^{\text{TE(TM)}}_{k_x,k_y}$ is the complex reflection coefficient of TE (TM) polarized light at the surface of the ground metal plane. For a normally incident light, Eq. (12) can be written as

$$\bar{\boldsymbol{\eta}}^S(0,0) = i\sqrt{\varepsilon^S/\mu^S}\tan\left(k_0 n_S d - i\ln(\bar{r}^{\text{TE}}_{0,0})/2\right) \quad (13)$$

Eq. 13 shows that the effective admittance of the combined dielectric layer and the ground plane is dispersive, which helps to cancel the intrinsic dispersion of the metal dipoles. Thus, with a proper thickness of the dielectric layer, a broadband and high-efficient optical device could be realized.

**III. Retrieval of Lorentzian parameters**

For an antenna of a specific geometry, the Lorentzian parameters can be retrieved by fitting the full wave simulation of a simple periodic array of the antennas with our analytical theory. Here CST microwave studio software is used to for the simulation. We consider a metasurface consisting of a periodic array of identical gold nanorods with the same orientation in a square lattice with period $\Delta$, as shown in Fig. 3 (a). The semi-infinite cover and substrate are air and SiO2 with refractive indexes of 1 and 1.45, respectively. The relative permittivity of gold is modeled by Drude model. That is

$$\varepsilon_r(\text{Au}) = 1 - \frac{\omega_p^2}{\omega(\omega + i\omega_\tau)}, \quad \omega_p = 1.37\times 10^{16}\,\text{rad/s}, \quad \omega_\tau = 1.215\times 10^{14}\,\text{rad/s}$$

For wavelength much greater than the period, we only need to consider the contribution of the zero-order Fourier component of the polarizability tensor. For a normally incident light from the cover side, combining Eq. (4), (6) - (8), the zero-order transmitted and reflected electric fields, denoted by subscript 0, can be written as

$$\begin{bmatrix} E_{0x}^{S+} \\ E_{0y}^{S+} \end{bmatrix} = \frac{2n_C}{\tilde{n}^2 + \left(k_0 \frac{\alpha_-}{\Delta^2}\right)^2} \begin{bmatrix} \tilde{n} + ik_0 \frac{\alpha_-}{\Delta^2} \cos 2\varphi & ik_0 \frac{\alpha_-}{\Delta^2} \sin 2\varphi \\ ik_0 \frac{\alpha_-}{\Delta^2} \sin 2\varphi & \tilde{n} - ik_0 \frac{\alpha_-}{\Delta^2} \cos 2\varphi \end{bmatrix} \begin{bmatrix} E_{0x}^{C+} \\ E_{0y}^{C+} \end{bmatrix}; \quad \boldsymbol{E}_{\perp 0}^{C-} = \boldsymbol{E}_{\perp 0}^{S+} - \boldsymbol{E}_{\perp 0}^{C+} \quad (14)$$

where $\tilde{n} = \left(n_C + n_S - ik_0 \alpha_+ / \Delta^2\right)$, $\alpha_\pm$ and $\varphi$ are defined in Eq. (4). For the periodic-structured metasurface, all the meta-atom are identical. Consequently, the subscript *mn* shown in Eq. (4) is omitted. Rewriting Eq. (14) in the circularly polarized bases, we have

$$\begin{bmatrix} \boldsymbol{E}_{0R}^{S+} \\ \boldsymbol{E}_{0L}^{S+} \end{bmatrix} = \frac{2n_C}{\tilde{n}^2 + \left(k_0 \frac{\alpha_-}{\Delta^2}\right)^2} \begin{bmatrix} \tilde{n} & ik_0 \frac{\alpha_-}{\Delta^2} e^{-i2\varphi} \\ ik_0 \frac{\alpha_-}{\Delta^2} e^{i2\varphi} & \tilde{n} \end{bmatrix} \begin{bmatrix} \boldsymbol{E}_{0R}^{C+} \\ \boldsymbol{E}_{0L}^{C+} \end{bmatrix} \quad (15)$$

Eq. (15) shows that only cross-circular polarizations in transmitted light can gain different additional phases for the normally incident beam with circular polarizations. The additional phases are non-dispersive geometric phases determined by the orientation angle of the meta-atoms.

Fig. 3 (b-e) shows the transmission amplitude and phase of the X (Y) -polarized light normally incident from cover onto the metasurface that is schematically shown in Fig. 3 (a). We fit the curves based on Eq. (5) and (14). The retrieved parameters of the nanorod are

$$g^S = 13.9\Delta^2, \quad \omega^S = 1.492 \times 10^{15} \text{ rad/s}, \quad \gamma^S = 6.0 \times 10^{13} \text{ rad/s}$$

$$g^P = 5\Delta^2, \quad \omega^P = 3.5 \times 10^{15} \text{ rad/s}, \quad \gamma^P = 3 \times 10^{13} \text{ rad/s}$$

The fitting results agree very well with the CST simulation. These fitting parameters will be used to model more complex metasurfaces consisting of antennas of spatially varying orientations for controlling the wavefront of light in the subsequent sections.

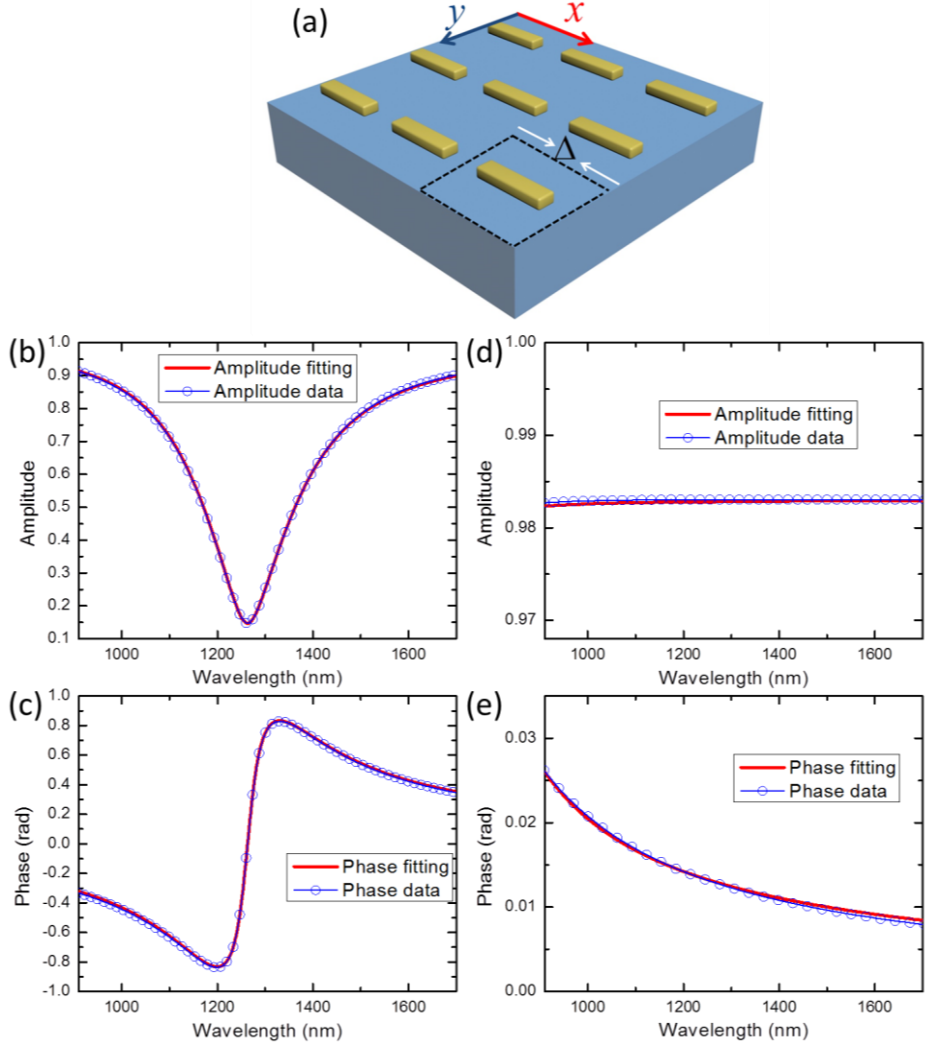

**Figure 3 (a)** Schematic illustration of a metasurface with periodic Au nanorods array on the SiO$_2$ substrate with period $\Delta$ is 500 nm. The width, length and thickness of the nanorod is 80 nm, 320 nm, and 30 nm, respectively. Numerical simulations of the transmission amplitude (b,d) and phase (c,e) of X-polarized and Y-polarized light normally incident from cover onto the metasurface. The fitting is based on Eq.(5) and (14).

**IV. Verification of the Analytical Theory with full-wave numerical simulations**

By feeding the retrieved parameters obtained from Section III into the analytical model, we can analytically model metasurfaces consisting of antennas of arbitrary orientations for manipulating the wavefront of incident beam. The analytical modeling results are compared to full wave simulations for several representative metasurfaces.

**A. Phase gradient metasurface**

We first consider a metasurface with one-dimensional geometric-phase gradient. In a periodic unit cell, the phase gradient metasurface comprises $N$ identical meta-atoms, whose orientations angle $\varphi$ exhibits a constant gradient $\pi/N$ along one direction. Fig. 4 (a) shows the simulated phase gradient metasurface with $N$ equal to 8, made up of the same Au nanorods as that in Fig.3 (a). The periods of the array along X axis and Y axis are $8\Delta$ and $\Delta$, respectively. From Eq.(4) and (6), we can obtain an approximate expression of the susceptibility tensor of the phase gradient metasurface by only keeping the first order spatial harmonic terms. That is

$$\boldsymbol{\chi} = \sum_{k_x,k_y} e^{ik_x x + ik_y y}\overline{\boldsymbol{a}}(k_x,k_y) \approx \frac{\alpha^S + \alpha^P}{2\Delta^2}\boldsymbol{I} + e^{i\frac{2\pi}{N\Delta}x}e^{2i\varphi_0}\frac{\alpha^S - \alpha^P}{4\Delta^2}\begin{bmatrix}1 & -i \\ -i & -1\end{bmatrix} + e^{-i\frac{2\pi}{N\Delta}x}e^{-2i\varphi_0}\frac{\alpha^S - \alpha^P}{4\Delta^2}\begin{bmatrix}1 & i \\ i & -1\end{bmatrix}$$

(16)

where $\varphi_0$ is the orientations angle of the first nanorod chosen in the periodic unit cell.

In Eq. (16), we only consider the -1, 0 and +1 order diffractive light for the phase gradient metasurface, whose wave vectors $(k_x, k_y)$ are given by $(k_x^{in} - 2\pi/N\Delta, k_y^{in})$, $(k_x^{in}, k_y^{in})$ and $(k_x^{in} + 2\pi/N\Delta, k_y^{in})$ for an incident light with wave vectors $(k_x^{in}, k_y^{in})$. Combining Eq.(6)-(9) and (15), we can obtain the expressions of the transmitted and reflected electric fields for the incident light from the cover,

$$\boldsymbol{E}_{\perp -1}^{C-} = \boldsymbol{E}_{\perp -1}^{S+} = (-2i\omega\varepsilon_0)\Lambda_{-1}^{-1}\overline{\boldsymbol{a}}_{-1}\boldsymbol{\Xi}^{-1}\boldsymbol{\eta}_0^C \boldsymbol{E}_{\perp 0}^{C+} \tag{17a}$$

$$\boldsymbol{E}_{\perp 0}^{S+} = 2\boldsymbol{\Xi}^{-1}\boldsymbol{\eta}_0^C \boldsymbol{E}_{\perp 0}^{C+}; \qquad \boldsymbol{E}_{\perp 0}^{C-} = (2\boldsymbol{\Xi}^{-1}\boldsymbol{\eta}_0^C - \boldsymbol{I})\boldsymbol{E}_{\perp 0}^{C+} \tag{17b}$$

$$\boldsymbol{E}_{\perp 1}^{C-} = \boldsymbol{E}_{\perp 1}^{S+} = (-2i\omega\varepsilon_0)\Lambda_{+1}^{-1}\overline{\boldsymbol{a}}_{+1}\boldsymbol{\Xi}^{-1}\boldsymbol{\eta}_0^C \boldsymbol{E}_{\perp 0}^{C+} \tag{17c}$$

where subscripts -1, 0 and +1 denote diffraction orders of the transmitted and reflected electric fields from the metasurface, respectively, and

$$\boldsymbol{\eta}_{\pm 1}^{S/C} = \boldsymbol{\eta}^{S/C}\left(k_x^{in} \pm \frac{2\pi}{N\Delta}, k_y^{in}\right), \quad \boldsymbol{\eta}_0^{S/C} = \boldsymbol{\eta}^{S/C}(k_x^{in}, k_y^{in}); \quad \overline{\boldsymbol{a}}_{\pm 1} = \overline{\boldsymbol{a}}\left(\pm\frac{2\pi}{N\Delta}, 0\right), \quad \overline{\boldsymbol{a}}_0 = \overline{\boldsymbol{a}}(0,0)$$

$$\Lambda_{\pm 1} = (\boldsymbol{\eta}_{\pm 1}^S + \boldsymbol{\eta}_{\pm 1}^C + i\omega\varepsilon_0\overline{\boldsymbol{a}}_0); \quad \Lambda_0 = (\boldsymbol{\eta}_0^S + \boldsymbol{\eta}_0^C + i\omega\varepsilon_0\overline{\boldsymbol{a}}_0)$$

$$\boldsymbol{\Xi} = \Lambda_0 + (\omega\varepsilon_0)^2 \overline{\boldsymbol{a}}_{+1}\Lambda_{-1}^{-1}\overline{\boldsymbol{a}}_{-1} + (\omega\varepsilon_0)^2 \overline{\boldsymbol{a}}_{-1}\Lambda_{+1}^{-1}\overline{\boldsymbol{a}}_{+1}$$

For a right circularly polarized light normally incident on the metasurface shown on Fig.4 (a) from the cover, using the retrieved parameters of the nanorod, we can obtain the reflectivity and transmittivity of the zero-order and first-order diffractive light from Eq.17 (a- c). Fig 4 (b,c) show that the analytical results agree well with those obtained from the CST simulation.

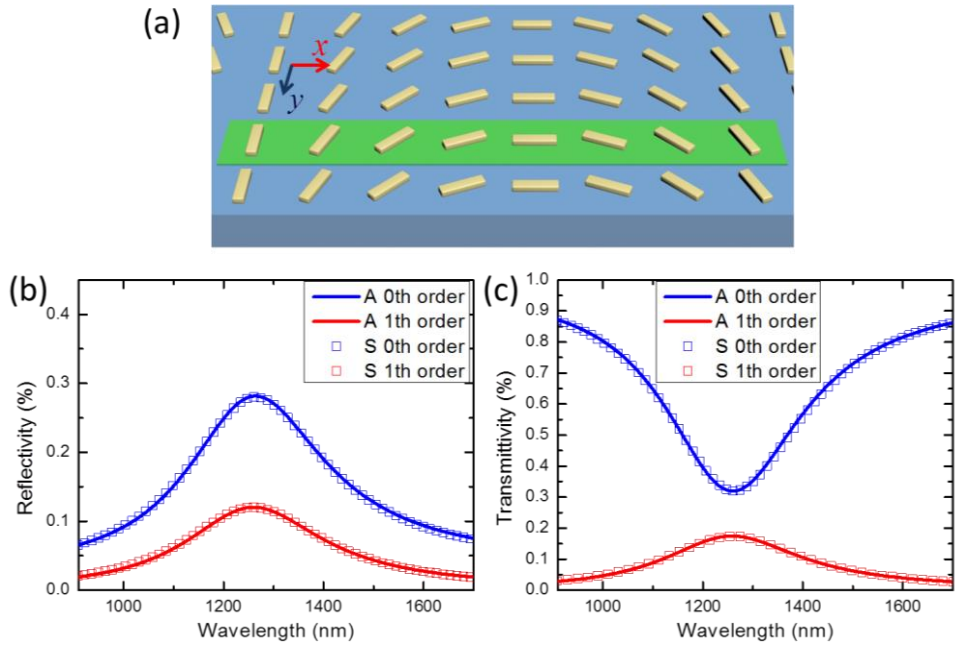

**Figure 4**. (a) Schematic illustration of a phase-gradient metasurface with a linear gradient in the orientations angle along the x axis. The nanorad is same as that in Fig. 3 (a). (b, c) The analytical and simulated reflectivity spectra (b) and transmittivity spectra (c) for both zero-order and first-order diffractions when a right circularly polarized light (RCP) light is normally incident from air onto the metasurface. In the insets, letter A and letter S denote results obtained from our analytical theory and CST simulation, respectively.

Furthermore, to check the applicability of Eq. 17(a - c) for the obliquely incident light, we simulate a RCP with an incident angle of $30^o$ from air onto the metasurface as shown in Fig.4 (a). Figure 5 shows that the analytical results still agree well with those calculated from CST simulation for the obliquely incident beam.

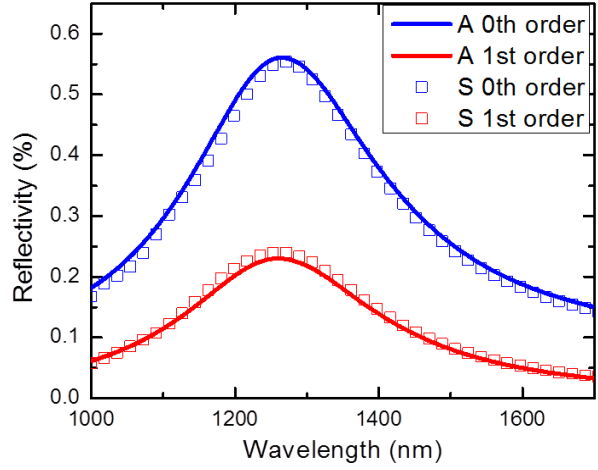

**Figure 5**. The analytical and simulated reflectivity spectra of zero-order (blue curve) and first-order (red curve) for a RCP beam at an incident angle of 30° from air onto the metasurface shown in Fig. 4(a).

**B. Reflective metasurface**

The configuration of the reflective metasurface is shown in Fig. 6 (a), which consists of an array of antennas atop a homogeneous dielectric layer and a ground plane. For simplicity, we confine ourselves to the discussion of incident light at normal direction. The admittance tensor of the combined dielectric layer and the ground plane is given by Eq.(13). For Au ground plane, that is

$$\bar{\boldsymbol{\eta}}^S(0,0) = -\boldsymbol{\eta}^S(0,0) i \tan\left(k_0 n_S d - i \ln(\bar{r}_{0,0}^{Au})/2\right) \tag{18}$$

Where, $\bar{r}_{0,0}^{Au}$ is the reflection coefficient of the Au plate. For a reflective metasurface consisting of antennas with identical orientation, we can obtain the zero-order reflected electric fields by replacing the refractive index of the substrate $n_S$ with $-i n_S \tan\left(k_0 n_S d - i \ln(\bar{r}_{0,0}^{TE})/2\right)$ in Eq. 14. Fig. 6 (b) shows the analytical and simulated reflectivity spectra of co-polarization and cross polarization for normally incident RCP onto the three layer metasurface. It is obvious that the metasurface can work as a reflective half-wave plate with the wavelength from 1100nm to 1400nm.

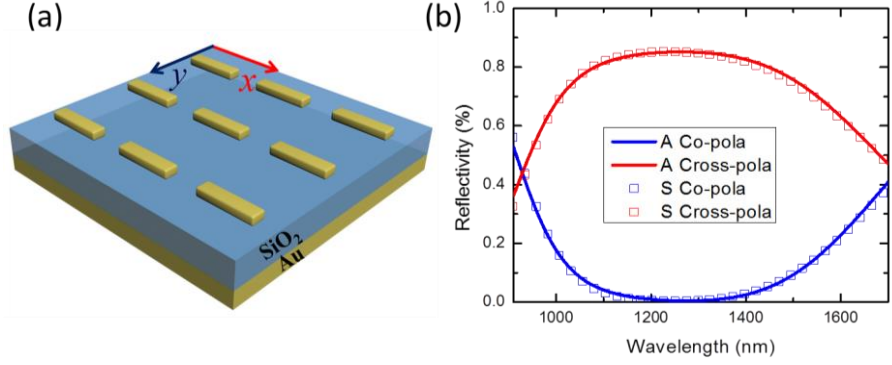

**Figure 6** (a) Schematic illustration of a three layer metasurface design with periodic Au nanorods array on the SiO₂ substrate with Au ground plane underneath. The nanorad is the same as that in Fig. 3 (a) and the thickness of the SiO₂ is 182nm. (b) The corresponding analytical and simulated reflectivity spectra of co-polarization and cross polarization for a normally incident RCP beam.

Next, we consider a reflective metasurface (shown in Fig.7 (a)) with a linear phase gradient. By replacing $\boldsymbol{\eta}^S$ with $\bar{\boldsymbol{\eta}}^S$ in Eq. 17, the analytical expressions of zero-order and first-order reflective electric fields from the metasurface can be obtained. Fig. 7 (b) shows the zero-order and first-order reflectivity obtained from both the analytical theory and CST simulation for a right circularly polarized light normally incident onto the phase-gradient metasurface. Again, the analytical results agree well with the simulated results.

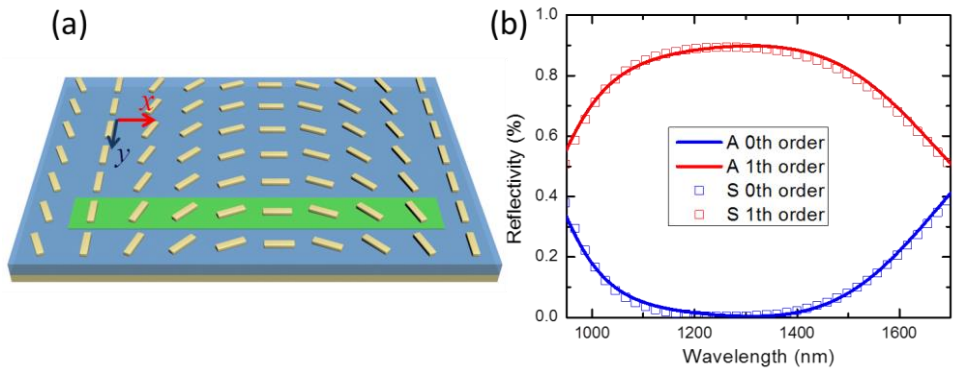

**Figure 7**. (a) Schematic illustration of a phase-gradient three layer design with an increased orientations angle along the x axis. The nanorad is same as that in Fig. 3 (a) and the thickness of the SiO₂ is 182nm. (b) The corresponding analytical and simulated reflectivity curves of zero-order and first-order diffractive light for the normally incident RCP.

### C. One-dimensional phase modulated metasurface

In order to further validate the applicability of the analytical theory for more complex phase profiles, we apply it to a metasurface with a carefully designed nonlinear phase gradient for selectively exciting a number of diffraction orders. The metasurface unit cell consists of 13 antennas with orientation angles with respect to the x axis designed as [0.9π, 0.06π, 0.21π, 0.34π, 0.39π, 0.37π, 0.47π, 0.61π, 0.76π, 0.93π, 0.14π, 0.46π, 0.72π]. As shown in Fig. 8, three diffractive orders from -3 to -1 are selectively excited over the whole displayed wavelength range from 970nm to 1700nm. Meanwhile, the zeroth order is very low across the wavelength between 1200 nm and 1400 nm. Both numerical simulation and analytical modeling are in good agreement. All the calculation and simulation prove the accuracy of the analytical theory, and it can be applied to the modeling of more complicated metasurface holograms.

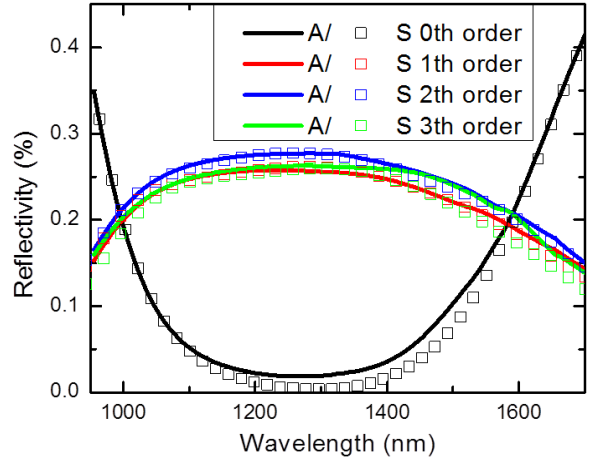

**Figure 8** The analytical and simulated reflectivity curves of different ordered diffractive light for the normally incident RCP onto the one-dimensional hologram metasurface ended with Au ground plate.

---

### V. Conclusion

In summary, we have derived a generally analytical theory of metasurfaces based on the assumptions that the tangential electric fields on both sides of the metasurface are continuous and the meta-atoms possess localized resonances with Lorentz-Drude forms. By retrieving the six parameters used to describe the anisotropic polarizability tensor of a meta-atom, optical performance of geometric metasurfaces constituted by the same meta-atoms but different orientations can be accurately calculated by the analytical formulae. As the analytical results show excellent agreement with full-wave numerical simulations, the analytical theory provides an efficient method to design and model optical devices based on metasurfaces of complex phase profiles.

This work was supported by EPSRC (Grant No. EP/J018473/1), National Nature Science Foundation of China (Grant Nos. 11374367, 61328503), and Leverhulme Trust (Grant No. RPG-2012-674).